\documentstyle[11pt]{article}          
\begin{document}
\title{{\huge \bf Impressionism and Surrealism in 
Multiparticle Dynamics}\thanks{Invited
talk at the CORR98 meeting on Correlations, M\'atrah\'aza,
June 1998, to appear in the proceedings of the meeting in
World Scientific. This talk is dedicated to the memory of Peter 
Carruthers.}}
\author{R. M. Weiner\thanks{E.Mail:
weiner@mailer.uni-marburg.de}
\date{Physics Department, University of Marburg,\\ Wieselacker
8, 35041 
Marburg, Germany \\and \\
Laboratoire de Physique Th\'eorique et Hautes 
\'Energies,
Univ. Paris-Sud,\\ 177 rue de Lourmel, 75015 Paris, France}}
\maketitle
\thispagestyle{empty}
\noindent
 
\newcommand{\lraw}{\leftrightarrow}
\newcommand{\mbold}[1]{\mbox{\boldmath $ #1 $}}
\newcommand{\be}{\begin{equation}}
\newcommand{\ee}{\end{equation}}
\newcommand{\benn}{\begin{displaymath}}		
\newcommand{\eenn}{\end{displaymath}}		
\newcommand{\ba}{\begin{eqnarray}}
\newcommand{\ea}{\end{eqnarray}}
\newcommand{\bann}{\begin{eqnarray*}}	
\newcommand{\eann}{\end{eqnarray*}}
\let\ul\underline
\renewcommand{\baselinestretch}{1.5}


\section*{Peter Carruthers and multiparticle dynamics; a
story of common interests} 

Peter Carruthers, the scientist, was a man of remarkable 
mathematical 
and physical culture. He had also something which is less
common now a days, namely a profound   
physical intuition which guided him to approach in an original way
the most interesting problems at a given moment and in 
this way he became a guide also for others. 

Multiparticle dynamics being by definition a many-body
problem, it has to use methods specific for many body-physics.
 A physicist like Peter
Carruthers, who had brought important and well known contributions to solid
state physics, was therefore best qualified to enter the
rapidly developping field of
multiparticle dynamics.
 It is however less known that he was
the first one to postulate the existence of a new state of
quark matter, which he called ``quarkium" and which he
assumed to be a ``bizarre Fermi liquid"\cite{Quarkium}. Note that this was
three years before the now accepted concept of deconfined
quark matter was proposed.

Two of the methods used in many-body physics are of
particular importance for multiparticle dynamics, 
hydrodynamics and quantum statistics and Pete brought
lasting contributions to both of these.   
His scientific achievements would certainly have been even
more numerous, hadn't he have to spend much of his 
time in the seventies and eighties with administrative
duties.
(He was the founding director of the theory division of 
the Los Alamos  
National Laboratory and later chairman of the Physics
Department of the University of Arizona.)

\paragraph{Local equilibrium and hydrodynamics}
 In the early seventies, when multiparticle production 
transcended cosmic rays and entered accelerator physics, it
became clear to some physicsts that ``conventional"
theoretical methods like Reggeology or current algebra 
were inadequate for the handling of the complex nature
of problems one was facing. When I had expressed this point of
view at the Batavia conference in 1972, I felt rather
isolated. I learned later from Pete, who 
apparently had
participated at that session, but whom I do not remember to
have met in
Batavia, that he had shared from the
beginning this insatisfaction. We were in good, but 
``restricted" company. I personally had had the opportunity
of many interesting discussions with Hagedorn during my
stay at CERN in 1969-1970, whose theory showed clearly how far 
``non-conventional" approaches could lead. 
I was also aware   
of course that E. L. Feinberg was a distinguished founding 
member of this ``club" , but this was (almost) it
\footnote{The remarkable contribution of Frautschi to the
Hagedorn bootstrap idea should also be mentioned.}. In
1972-1974 appeared the papers by Pete and Minh on the Landau
hydrodynamical model where they showed that many empirical
observations made in, at that time, ``high energy" p-p
reactions could be explained by this model. Most
instrumental was however the now famous paper by Pete
``Heretical Models of Particle Production" \cite{Heretical} 
in which he resumed 
the successes, up to that date, of the Landau model in multiparticle dynamics.
The very title of the paper is self-explanatory. 
 
The superiority of the Landau and the Hagedorn models as compared with all
other approaches to multiparticle dynamics consists, among
other things, in the fact that these are the only models which can
explain the limitation of transverse momenta of secondaries,
which is the most characteristic phenomenological 
property of strong interaction physics. 
Other models just postulate this property (in the
Regge model e.g. this is reflected in the empirical 
$\beta$ function). Furthermore, the Landau model {\em predicted}
the violation of Feynman scaling at high energies and 
 Pete and collaborators had a tough time in convincing
adepts of ``conventional" physics that the data showed that 
Landau was right and that they were wrong.
This story had a follow-up in heavy ion physics (cf. the end
of this section).

On the other hand there is apparently a
high price to be paid for the possibility to apply a
classical method like hydrodynamics to particle physics. It
is the assumption of local (thermodynamical) equilibrium
(LE) and  
the constraint that correlations between positions and 
momenta of particles should not exceed the limits imposed by
the Heisenberg incertitude relations. The justification 
of LE has preoccupied us for quite a time and lead to the
organization of the series of meetings\cite{LESIP} LESIP 
starting in
1984. It has found apparently an explanation
 in the realization
that the formation of hadronic matter is most probably 
preceded by the phase of quark matter. 

The classical versus quantum issue stimulated Pete to 
write with Zachariasen a review paper \cite{CZ} on the 
applicability of
the Wigner function to multiparticle dynamics, this function
being at the border between classical and quantum physics. 
(The topic of the Wigner function will be discussed below 
also in another context.) 

\paragraph{Correlations}
 Classical fields play also
a major role in the modern developments of particle physics 
through spontaneously broken symmetries and the
Goldstone-Higgs-Kibble mechanism. This lead Fowler and
myself to
investigate the ``Effects of Classical Fields in Meson
Correlations"
\cite{FW} out of which resulted a long series of studies on
correlations\footnote{LESIP IV was devoted almost
exclusively to the topic of correlations and got its title
CAMP (Correlations and Multiparticle Production) from this
 subject. It is gratifying to see that the series of meetings
on fluctuations started by Pete in Aspen in 1986. and 
which has continued 
under various names up to the present one, has adopted now the
title CORRELATIONS. This choice is quite appropiate as
fluctuations are essentially based on correlations.}    
and multiplicity distributions.  

On the issue of correlations Pete's and my interests again
met, since Pete (with Nieto) had contributed with important papers to the
subject of coherent states. This also explains how Pete was
among the first to understand why we thought that
Bose-Einstein correlations were at the center of interest of
multiparticle dynamics,
at a time when practitioners in this field were mostly concerned with
multiplicity distributions and had not yet realized that
multiplicity distributions and correlations were
complementary subjects. This explains also our collaboration
in \cite{int} which was the first paper to suggest that the
so called intermittency effect was essentially a mere
consequence of BEC. \footnote{This idea was presented by
Erwin Friedlander at LESIP III.}       

A follow-up of Pete's work on multiplicity distributions was
his interest in complexity studies. He was probably the first
one to give a course on this subject, a copy of which I was
happy to get from him in the early nineties.  

Peter Carruthers was not only an outstanding scientist, but
also a very talented poet and painter and some of his poems
appear published for the first time in this volume. 
He was a true Renaissance man interested in all what is
human and an admirer of European culture. 
However his artistic preoccupations never interfered
destructively with his scientific work. He 
always distinguished between rigour in science and artistic
freedom. 

While (some of) 
his paintings were impressionistic or surrealistic, his physics 
papers met always highest scientific standards, despite the 
fact that they were very imaginative. Unfortunately, in 
some of the recent literature on multiparticle dynamics and 
in particular in high energy heavy ion physics, 
these standards are not always respected
\footnote{That is why Pete and
others avoided lately ``Quark Matter" meetings.}.
          An example is the erroneous
``impression" of a large part of the heavy ion community about the
existence of the rapidity plateau. In the discussions
preceding the design of the ALICE detector for heavy ion reactions
at the CERN LHC it took great efforts to convince some
people to extend the accessible rapidity range beyond one
rapidity unit, as planned initially. The idealization of
boost invariance made such an extension appear superflous.
Other examples of ``impressionism" and ``surrealism" in physics, in 
particular in BEC will be given below\footnote{For an
introduction to the subject of BEC cf. a forthcoming
textbook by the author to be published by John Wiley \& Sons
in 1999. This will be quoted in the following as [I].}.  

\section*{Impressionism in Bose-Einstein correlations} 

\begin{flushright}{\footnotesize Impressionism:``... 
a theory and practice of presenting the most immediate
...aspects\\
...with ...little study of ... realistic detail"
(Webster's Comprehensive Dictionnary)}
 \end{flushright}

\paragraph{Uses and ab-uses of the Wigner function in BEC} 

 The experimental observation of the
fact that the two particle correlation function depends not
only on the difference of momenta $q=k_1-k_2$ but also on 
the sum $k_1+k_2$ lead to the introduction 
\cite{Pratt} of a ``source" 
function within the well known Wigner function formalism of
quantum mechanics. (For its relativistic generalization and 
application in hydrodynamics cf.
\cite{bernd1}). 
\footnote{An attempt \cite{Yano} to consider the
correlation between coordinates and momentum
 within the ordinary wave function 
formalism was shown in \cite{Axel} 
 to have 
pathological features as it
leads in some cases to a violation of the lower bounds of the
correlation function.}   
While it turned out later \cite{AW} that this
property of the correlation function can be derived within
the current formalism without the approximations involved by
the Wigner formalism, this formalism is still useful 
 within a hydrodynamical context. That is why we will
decribe this formalism in the following.

The Wigner function called also source function, 
$g(x,k)$, may be 
regarded as the quantum analogue of the density of particles
of momentum $k$ at space-time point $x$ in classical statistical
physics. It is defined within the wave function formalism as

\begin {eqnarray}
g({\bf x},{\bf k},t)& =&\int d^3x^{'}\psi^*\left({\bf
x}+\frac{1}{2}{\bf x^{'}},t\right)\psi\left({\bf x}-\frac{1}{2}{\bf
x^{'}},t\right)e^{i{\bf kx{'}}}\nonumber\\
&=&\int d^3k^{'}\psi^*\left({\bf k}+\frac{1}{2}{\bf k{'}},t\right)
\psi\left({\bf k}-\frac{1}{2}{\bf k{'}},t\right)e^{-i{\bf k{'}x}}
\label{eq:Wigner}
\end{eqnarray}  
and is related to the coordinate and momentum densities
 by the equations
\be
n({\bf x},t)=\int d^3kg({\bf x},{\bf k},t)
\label{eq:(B)}
\ee
and
\be
n({\bf k},t)=\int d^3xg({\bf x},{\bf k},t)
\label{eq:(C)}
\ee
respectively.
To be able to use it in the context of hydrodynamics
which is a classical theory and where it should approximate
as far as possible the Boltzmann distribution function, one
has to make sure that the 
quantum effects (i.e. the weight of the domain where $g(x,k)$ 
takes negative values) are small.
This is achieved by
expanding the wave function in powers of Planck's constant 
$\hbar$. It turns
out that the
semi-classical approximation is valid as long as the 
relation 
\be
\frac{\Lambda}{4\pi}\mid\frac{dk}{dx}\mid \ll k.
\label{eq:semicl}
\ee
is satisfied.
Here $\Lambda$ is the de Broglie wave
length of the particle. In the case of BEC we deal with
correlations of particles which originate from the entire
source. Eq.(\ref{eq:semicl}) then implies that the
quantum effects are small provided $k$ does not vary very
much across a region of the order of $L$ where $L$ is a
typical length characterizing the source. Using an analogy,
 one might express this
condition by saying that the gradients
of temperature across the ``fireball" should be small.
However even this condition is not sufficient if one considers
particles which are produced from the same space-time point.
 This implies that the ``surprising" effects, in particular
particle-antiparticle correlations \cite{APW1} 
cannot be treated with the conventional
Wigner function approach presented above. 
 
Furthermore it turns out that the Wigner
function is useful for BEC only if a more stringent
condition is fulfilled, namely that the difference of
momenta $q$ of the pair is small. It is thus clear that its
applicability  is more restricted
than that of the classical current approach, where only the
``no recoil" condition, i.e. small total momentum of produced
particles must be respected.   
This circumstance 
is often overlooked when comparing theoretical predictions 
based on the Wigner approach with experimental data. 
In particular herefrom also follows that the application of the
Wigner formalism to data has necessarily to take into account
from  the beginning resonances which dominate the small 
$q$ region. 
 Heuristically 
 the use of the Wigner function for BEC
is justified only in special cases as e.g. when a coherent hydrodynamical study is
performed, i.e. when the observables are related to an
equation of state and when simultaneously single and higher
order inclusive dstributions are investigated. 
Unfortunately only very few papers where the Wigner function
formalism is used are bona fide hydrodynamical studies. The
majority of  ``theoretical" papers in this context are
``pseudo-hydrodynamical" in the sense that the
source function $g$ is expressed in terms of {\em effective} physical variables
like effective temperature or effective velocity, which are not related by an
equation of state. The choice of the form of $g$
 is up to the {\em impression} of the practitioner. 
In this procedure 
 the application of the Wigner
approach is a luxury. This is a
fortiori
 true given the fact that
the Wigner approach is mathematically not simpler that the classical current
approach, of which it is a particular case. 

In second quantization
$g(x,k)$ is defined in terms of the correlator 
$< a^\dagger ( {\bf k}_i ) a ( {\bf k}_j )>$ 
by the relation
\begin{equation}
< a^\dagger ( {\bf k}_i ) a ( {\bf k}_j )  > =
\int d^4 x \exp [- i x_\mu ( k_i{}^\mu  - k_j{}^\mu ) ] \cdot
g [ x , \textstyle{\frac{1}{2}} ( k_i + k_j ) ]
\label{eq:wigdef}
\end{equation}
This is a natural generalization of eq.(\ref{eq:(C)}) to which
 it reduces in the limit $k_{i}=k_{j}$.

The relation between the Wigner approach and the classical
current approach is established by expressing the rhs of 
eq.(\ref{eq:wigdef}) in terms of the currents. One has
\begin{equation}
g(x,k)\ =\ \frac{1}{2 \sqrt{E_{i}E_{j}}(2\pi)^{3}} \ 
 \int d^4z \ <J\left(x+\frac{z}{2}\right) 
J\left(x-\frac{z}{2}\right)> \ \exp\left[-ik^\mu z_\mu\right]
\label{eq:wigcur}
\end{equation}
The derivation of the Wigner formalism from the classical
current formalism has the important advantage that it avoids
violations of quantum mechanical bounds as those mentioned
previously.

>From the above considerations follows that 
a parameter in the space-time approach, such as the 
correlation length $L$, can
be linked to parameters that enter the Wigner function, e.g.,
the temperature $T$ if the system is in thermal equilibrium. 
This particular case of a relation between $L$ and $T$ 
was derived in \cite{APW}.

As mentioned already, the use of the Wigner formalism
 is worthwhile within a true
hydrodynamical approach when the relation with the
equation of state is exploited. In this case 
the probability to produce a particle of momentum $k$
from the space-time point $x$ then depends on the fluid velocity, 
$u^\mu(x)$ and the temperature $T(x)$ at this point, and one
has  
\begin{eqnarray}
&&\sqrt{E_i E_j} < a^\dagger ( {\bf k}_i ) a ( {\bf k}_j ) >
 \nonumber\\
&&= \frac{1}{(2\pi)^3} \int\limits_\Sigma
\frac{\textstyle{\frac{1}{2}} (k_i{}^\mu + k_j{}^\mu)
d \sigma_\mu (x_\mu)}{\exp \left[\displaystyle{
\frac{\textstyle{\frac{1}{2}} (k_i{}^\mu + k_j{}^\mu)
u_\mu (x_\mu)}{T_f (x_\mu)} } \right] - 1
} \cdot \exp [- i x_\mu (k_i{}^\mu - k_j{}^\mu)] 
\label{eq:hydro}
\end{eqnarray}
Here, $d\sigma^\mu$ is the volume element on the 
freeze-out hypersurface $\Sigma$ where the final state 
particles are produced.
Despite the fact that the use of the Wigner formalism can
be defended only if combined with bona-fide hydrodynamics and an
equation of state\footnote{That full-fledged hydrodynamics
is sometimes indispensable was illustrated e.g.
in the study of the role of the transverse and longitudinal
expansion on the effective radii \cite{bernd1}.}, 
with the exception of a few real, 
albeit numerical, hydrodynamical 
calculations, most phenomenological papers on BEC in heavy ion
reactions
have used the Wigner formalism without a proper
hydrodynamical treatment, i.e. without solving the equations
of hydrodynamics; hydrodynamical concepts
like velocity and temperature were used just to parametrize the
Wigner source function. While such a procedure may be 
acceptable as a theoretical exercise, it is certainly no
substitute for a professional analysis of heavy ion
reactions. This is particularly true when real data have to be
interpreted\footnote{A recent experimental paper \cite{2Na49}
where such
an analysis is performed is a good illustration of the
limits of
 pseudo-hydrodynamical models. Despite the fact that the
statistics are so rich that ``the statistical errors on the
correlation functions are negligible", the outcome of the
analysis is merely the resolution of the ambiguity between
temperature and transverse expansion velocity of the source.  
It is clear that such an
ambiguity is specific to pseudo-hydrodynamics and is from
the beginning absent in a correct hydrodynamical treatment. 
Moreover even this result is questionable given some
 doubtful assumptions
which underly this analysis. To quote just two: 
(i) The assumption of boost
invariance made in \cite{2Na49} decouples the longitudinal 
expansion from the
transverse one. This not only affects the conclusions drawn 
in this analysis but prevents the (simultaneous) 
interpretation of the
experimental rapidity distribution.
(ii) The neglect of long lived resonances which strongly
influences the $\lambda$ factor (cf. eq.(\ref{eq:C2}) below)
and thus also the extracted
radii. 
 Of course, despite the claimed richness of the data, no
attempt to relate the observations to an equation of state
 can be made within this naive phenomenological approach.}.  
 
As examplified in \cite{bernd1} such a
procedure is unsatisfactory, among other things because it
can lead to wrong results.

The use of this ``pseudo-hydrodynamical"
approach is even
more surprising if one realizes that the Wigner
formalism not only is not simpler 
than the more general current
formalism but it is also less economical. The number of independent 
parameters necessary to characterize the BEC within the
Wigner formalism is (cf. e.g. ref.\cite{Nix}) 
$10$, i.e. it is as 
large as that in the current formalism \cite{APW}. However 
the $10$
parameters of \cite{Nix} describe a very particular source
\footnote{To consider such an approach as ``model 
independent" as has become customary in the heavy ion
literature is misleading.},
as compared with that of the current formalism: besides the
fact that the second order correlation function $C_2$ in the
Wigner approach is assumed to be
Gaussian, it is completely chaotic and it can provide
only a correlation length $L$. 
In the current formalism on the other hand, with $10$
parameters the correlation function is not restricted to the
Gaussian form, one describes also coherence and one
distinguishes between the correlation length $L$ and the
geometrical radius $R$.
For the search of quark gluon plasma,  
 $R$ is the relevant quantity, because the energy density is
defined in terms of $R$ and the use
of $L$ instead of $R$ leads to an overestimate of this
energy density.

Furthermore the physical significance of the parameters
of the Wigner source is unclear if the Gaussian assumption does not hold. 
Not only is there no apriori reason for a Gaussian 
form,  but on the contrary, 
 both in particle physics and
in quantum optics, there exists experimental evidence that
in many cases an exponential function in $|q|$  is at small
$q$ a better
approximation for $C_2$ than a Gaussian. Furthermore
, it is known \cite{Weiner} that in the
presence of coherence, no single simple analytical 
function, and in particular no single Gaussian is expected 
to describe $C_2$. This is a straightforward consequence of 
quantum statistics. Last but not least, in heavy ion
reactions it has been shown that resonances deform any
Gaussian form into a non-Gaussian one. 

Given the fact that good experimental BEC data are expensive
both in terms of accelerator running time and man-power, 
the use of inappropiate theoretical tools, when more
appropiate ones are available, is a waste which has
to be avoided.

\subsection*{Higher order correlations}

The modern treatment of BEC is based on the density matrix
$\rho$ and field theory. In quantum optics (QO) one writes 
$\rho$ in terms of the coherent state representation and 
the most frequently form for this expansion is the Gaussian
one\footnote{The Gaussian form of this representation must
not be confused with the form of the correlations function.}.

Besides its mathematical convenience this form follows from
the central limit theorem for an infinite number of
independent fields. One most remarkable consequence of the
Gaussian form of the density matrix is the
fact that all higher order correlation functions are determined just
by the first two correlation functions. However this does
not imply at all that higher order correlation measurements are
unnecessary, once the first two correlation
functions are determined. Indeed there are 
  at least three reasons why the measurement of higher order
correlations is important:

(i) In the absence of
a theory which determines from first principles the
first two correlation functions,
 models for these quantities are used,
which are only approximations. The errors introduced by
these phenomenological parametrizations manifest themselves
differently in each order and thus violate the central 
limit theorem.

(ii) In experiments, because of limited statistics and
sometimes also because of theoretical biases not {\em all}
physical observables are determined, but rather 
averages over certain variables are performed, which again
introduces errors which propagate (and are amplified) from
lower to higher correlations.
  
(iii) The conditions of the applicability of the above theorem
and in particular the postulate that the number of fields 
(sources)
is infinite and that they act independently can never be
fulfilled exactly. 

Conversely, from the comparison of correlation functions of
different order one can test the applicability of the
central limit theorem and pin down more precisely the 
 parameters which determine the first two correlation
functions, which 
is essentially the 
purpose of particle interferometry.

Initially theoretical calculations for higher order BEC were
based on this QO formalism.        
After a slight confusion related to the use of an
incorrect formula for the comparison of the various orders
of correlation functions with data \cite{UA1}, it was found
\cite{mad} that in a first approximation a simplified
formalism of this type using a single BEC variable, 
the invariant
momentum difference $Q$, could account 
for the experimental measurements. For a discussion of these
issues cf. also \cite{Coll}.
  
In the mean time further theoretical and experimental
developments took place. 

On the experimental side a new technique 
for the study of higher order correlations
was developped (this method was developped in great part by
collaborators of Peter Carruthers),
the method of correlation integrals which was applied \cite{
UA1New} to a subset of the same UA1 data in order to test the
above quoted QO formalism. The fits were
restricted to second and third order cumulants only.
As in \cite{UA1} it was found that by extracting the 
effective parameters (chaoticity and radius) 
 from the second order data, the ``predicted" third
order correlation, this time by using a correct QO formula, 
differed significantly from the measured one. 
If confirmed, such a result could indicate that the QO
formalism provides only a rough description of the data and
that higher precision data demand also more more realistic
theoretical tools.
\footnote{A further, but more remote possibility would be to look for 
deviations from the Gaussian form of the density matrix.
However, a more mundane reason for the result of ref.\cite
{UA1New} 
will be mentioned below.}.
Such tools are the quantum statistical
(QS)
 space-time approach
to BEC developped in the nineties in \cite{APW} and
mentioned above. 
This approach 
based on the classical current formalism is more 
appropiate to particle physics than
the quantum optical approach and 
 is a generalization of the former. The fields are replaced
by currents and 
 the parametrization of the
space-time distribution of the source, which is introduced 
in the
the QO formalism essentially only   
through dimensional considerations
, gets a more rigorous
foundation. Moreover, and most importantly, it turns out 
that there are two different length scales in BEC, one
related to the space-time distribution of the source and 
another associated with the
correlator and that $Q$ is not a natural variable for BEC
if one wants to obtain from it information about the
geometry of the source, the correlation length, and the chaoticity.
To get this type of information from measurements made in the variable
$Q$ a complex projection involving integrations over
unobserved physical quantities has to be performed
\footnote{This topic was discussed in detail in
\cite{intpr} in connection with 
the issue of intermittency, after it had been suggested by 
Bialas \cite{bi} that the
source itself may not have a fixed size, but rather a size 
which
fluctuates from event to event with a power distribution.
 In \cite{intpr} it was proven that,
 by starting from a space-time correlator with a
{\em fixed} correlation length and a source distribution with a
{\em fixed} radius, one gets after integrations over the
unobserved variables, a correlation function which mimics
power-behaviour.}.  

However this new approach, although more advanced, 
does not make redundant the
determination of higher order correlations as 
the considerations (i)-(iii) continue to be valid.

The space-time formalism 
 has been used recently in \cite{Nelly} 
to study higher
order correlations up to the fourth order in the variable $Q$ and the calculated
results were compared with the Na22 data \cite{Na22}. It was
found that the data could be fitted without difficulties 
with quite reasonable space-time parameters. At the same
time it was found that a possible reason for the negative
results obtained in \cite{UA1New} within the simpler QO
approach was the questionable
procedure used for testing the relation between second and
third order correlations. 
 Indeed
in \cite{UA1New} one did not perform a {\em simultaneous} fit of
second and third order data to check the QO formalism.
Such a simultaneous fit is necessary before drawing
conclusions, because
as
mentioned above (cf.(i) and (ii)), the errors involved in
``guessing" the form of the correlator, and the fact that 
the variable $Q$ does not characterize completely the
two-particle correlation, limit the applicability of the
theorem which reduces higher order correlations to first and
second ones. As a matter of fact, it was found \cite{Nelly} 
 that the second order  
correlation data are quite insensitive to the values of the
parameters which enter the correlator, while once higher 
order data are used in a simultaneous fit, a
strong delimitation of the acceptable parameter values
results. Thus there are several possible solutions if one
restricts the fit to the second order correlation and the
correct one among these can be found 
only by fitting {\em simultaneously} all correlations. If by
accident one chooses in a lower correlation the wrong
parameter set, then the higher correlations cannot be fitted
anymore.

\subsection*{Photon versus hadron interferometry}

Photons and mesons are both bosons and therefore satisfy the
same Bose-Einstein statistics. This leads to 
similarities in the corresponding Bose-Einstein correlations
which underly photon and hadron intensity interferometry.
However there are also differences 
between the two effects and some of these will be analyzed in the 
following.

Photon interferometry in particles physics is from a 
certain point of view superior to 
 hadron interferometry, because photons are weakly
interacting particles, while hadrons interact strongly.
This has two important consequences in photon BEC: (i) there
is (up to higher order corrections) no final state
interaction between photons, so that the BEC effect is
``clean"; (ii) in a high energy reaction, hadrons are
produced only at the end of the reaction (at
freeze-out), while photons are produced from the beginning, so that
photons can provide unique information about the initial
state. For the search of quark-gluon plasma this is
essential, because if such a state of matter is formed, then
this happens only in the early stages of the reaction. 
This is also important in lower energy heavy ion reactions
where the dynamics of the reaction as well as its space-time 
geometry are studied in this way.

These advantages of photon interferometry have stimulated
theoretical and experimental studies, despite the technical
difficulties due to the small rates of photon production and
the background due to $\pi^0$ decays. 

Besides the difference in the coupling constant, photons and
hadrons (for the sake of concreteness we shall refer in the
following to pions) have also other distinguishing
properties like spin, isospin, and mass which manifest
themselves in the corresponding BEC
and which sometimes are overlooked. The role of spin
 will be discussed below
\footnote{For a more detailed comparison of photon and
hadron interferometry cf.[I].}.

\paragraph{Photon spin and  
bounds of BEC.}

In refs. \cite{Neu} and \cite{Leo} it was found that
while for (pseudo-)scalar pions the intercept of the second
order correlation function $C_2(k,k)$ is a constant,
 even for unpolarized photons the intercept is a function of k.
One thus finds that, while for a system of charged pions (i.e. a mixture of $50\%$ 
positive and $50\%$ negative) the maximum value of the 
intercept Max$C_2(k,k)$ is 1.5, for
photons  Max$C_2(k,k)$ exceeds this value and this excess 
reflects the space-time properties of the source,
 the degree of (an)isotropy of the source, and the supplimentary
degree of freedom represented by the photon spin. 
The fact that 
the differences between charged pions and photons
 are enhanced for soft photons reminds us of a similar
effect found with neutral pions (cf. ref.\cite{APW}). 
Neutral pions are in general more bunched than identically 
charged ones and this difference is more pronounced for soft
pions. This similarity is not accidental, because photons 
as well as
$\pi^0$ particles are neutral and this circumstance has
 quantum field theoretical implications which will be
mentioned also below.

 The results quoted above, in particular those obtained by
Neuhauser \cite{Neu} were
challenged by Slotta and Heinz \cite{Heinz}. 
Among other things, these authors claim that for photon
correlations due to a chaotic
source ``the only change relative to 2-pion interferometry
is a statistical factor $\frac{1}{2}$ for the overall
strength of the correlation which results from the
experimental averaging over the photon spin". In \cite{Heinz}
a {\em constant} intercept $\frac{3}{2}$ is derived which is 
in contradiction with the results presented above.

We would like to point out here that the reason for
the difference between the results of \cite{Neu},\cite{Leo}
on the one hand and those of ref.\cite{Heinz} on the other
is mainly due to the fact that in 
\cite{Heinz} a formalism was used which is less general
than that used in \cite{Neu} and \cite{Leo} and which is 
inadequate for the present problem. This implies among 
other things that unpolarized photons cannot be treated in
the naive way proposed in \cite{Heinz} and that the results
of \cite{Neu} and \cite{Leo} are correct, while the results
of \cite{Heinz} are not.

 In 
\cite{Heinz} 
the following formula for the second order correlation 
function is used: 
\ba
C_2({\bf k}_1,{\bf k}_2)=
1+\frac{\tilde {g}_{\mu \nu}({\bf q,K})\tilde {g}^{\nu \mu}({\bf -q,K})}
{\tilde {g}^{\mu}_{\mu}({\bf 0,k}_1)\tilde {g}^{\mu}_{\mu}(
{\bf 0,k}_2)}
\label{eq:heinz23}
\ea

Here $\tilde{g}$ is the Fourier transform of a source 
function $(g(x,K)$ and ${\bf q}={\bf k}_1-
{\bf k}_2$, ${\bf K}=\frac{1}{2}({\bf k}_1+{\bf k}_2)$. 

This formula is a particular case of a more general formula
for the second order correlation function derived
 by Shuryak \cite{Shuryak}
using a model of uncorrelated sources, when 
  emission of particles from the same space-time point is
negligible. 

As is clear from this derivation
there exists also a third term, neglected in
eq.(\ref{eq:heinz23}) and which corresponds to the 
simultaneous emission of two particles from the same point
 (cf. \cite{APW}). 
While for massive
particles this term is in general suppressed, this is not
true for massless particles and in particular for soft
photons. Indeed in \cite{Neu} and \cite{Leo} this additional
term had not been neglected as it was done subsequently in 
\cite{Heinz}   
and therefore it is not surprising that
ref.\cite{Heinz} could not recover the results of
refs.\cite{Neu} and \cite{Leo}. The neglect of the term
corresponding to emission of two particles from the same
space-time point 
 is not permitted in the present case.
Emission of particles from the same space-time point  
corresponds in a first approximation to
particle-antiparticle correlations and this type of effect
leads also to the difference between BEC for identical 
charged pions and the BEC for neutral pions.
This is so because neutral particles coincide with the
corresponding antiparticles. (As a consequence of this
circumstance e.g.
while for charged pions the maximum of the intercept is 2, 
for neutral pions
it is 3 (cf. \cite{APW}). Photons being neutral particles,
similar effects like those observed for $\pi^0$-s are
expected and indeed found.

 This misapplication of the current formalism  
 invalidates completely the conclusions of 
ref.\cite{Heinz}. 

  Intuitively the fact that for unpolarized
photons  Max$C_2(k,k)$ is 2 and not 1.5 as stated in
\cite{Heinz}, can be explained as follows: a system of 
unpolarized photons consists on the average of $50\%$
photons with the same helicities and $50\%$ photons with
opposite helicities. The first ones contribute to the
maximum intercept 
with a factor of $3$ and the last ones with a factor of 1
(coresponding to unidentical particles).

For further details of the topics discussed here cf.
\cite{RW}.

\section*{Surrealism and pion condensates; pasers?} 

\begin{flushright}{\footnotesize Surrealism: ``A movement
...characterized by the incongruous\\ 
and startling arrangement and presentation of subject matter"\\ 
(Webster's Comprehensive Dictionnary)}
 \end{flushright}

 One of the most important 
effects of quantum optics which is
based on 
coherence is the phenomenon of {\em lasing}. Lasers are
 Bose condensates and it has been
speculated that such condensates, in particular pion
condensates, may exist also in nuclei
(cf. e.g. \cite{Migdal}) or be created in heavy ion reactions
(cf. e.g. \cite{Pratt1}, \cite{Ornik}). 

 However there exist important 
differences between photon condensates i.e. lasers and pion
condensates. Furthermore there are different theoretical
approaches to the problem of pion condensates and 
some confusing statements
 as to how pion condensates are produced. 
In the following we shall discuss briefly
 these issues. 

\paragraph{The multiplicity dependence of BEC} 

BEC for inclusive processes, which constitute by far the
most interesting and most studied reactions both 
with hadrons and photons, have to be treated by quantum field
theory, which is the appropiate formalism when the number of
particles is not conserved. For certain purposes however, sometimes one is interested in considering events with a fixed
number of particles. Thus the number of particles in a given event can help
selecting central collisons with small impact parameter.
Theoretically this situation can be handled within field
theory, using the methods of quantum statistics \cite{Shih}.
On the other hand for the construction of event generators wave
functions appear so far to be a convenient tool and
therefore, and also for historical reasons,
 some
theorists have continued to use the ``traditional" method of
wave function (wf). This implies
the explicit symmetrization of the products of single
particle wf, while in field theory the
symmetrization (of amplitudes) is authomatically achieved through the
commutation relations of the field operators. When the
multiplicities are large, the symmetrization of the wf
 becomes tedious.
This lead  
 Zajc \cite{Zajc}  
 to use numerical Monte
Carlo techniques for estimating $n$ particle symmetrized 
probabibilities, which he then applied to calculate
  two-particle BEC. He was thus able also to study the
question of the dependence of BEC parameters 
on the multiplicity $n$.
Using as input a second order BEC function parametrized in the form
\be
C_2 \sim 1+ \lambda \exp(-{\bf q}^{2} {\bf R}^2),  
\label{eq:C2}
\ee
where ${\bf q}$ is the momentum transfer and ${\bf R}$ the
radius, Zajc found, 
and this was confirmed in \cite{Shih}, that the
``incoherence" parameter
$\lambda$ decreased with increasing $n$ \footnote{In
\cite{Zajc} the clumping in phase space due to Bose symmetry
was also illustrated.}.
   
However Zajc did not consider that this
effect means that events with higher pion
multiplicities are denser and more coherent. On the
contrary he warned against such an 
interpretation  and concluded that his results
have to be used in order to eliminate the {\em bias} introduced by
this effect into experimental observations. 
\footnote{The same
interpretation of the multiplicity dependence of BEC was
given in \cite{Shih}. In this reference the nature of the 
``fake"
coherence induced by fixing the multiplicity is even
clearer, as one studies there explicitely partial coherence 
in a consistent, quantum statistical formalism.}

This warning apparently did not deter the authors of \cite{Pratt1} and 
\cite{Chao} to do just that. Ref.\cite{Pratt1} went even so far
to derive the possible existence of pionic lasers (pasers)
 from 
considerations of this type.

In a concrete example Pratt considers
a non-relativistic source distribution $g$ in the absence of
symmetrization effects:

\be
g(k,x)=\frac{1}{(2\pi R^2mT)^{3/2}}\exp\left(-\frac{k_0}{T}-
\frac{x^2}{2R^2}\right)\delta(x_0)
\label{eq:Pratt8}
\ee
where 
\ba
k_0/T=k^2/2{\Delta}^2
\ea

Here $T$ is an effective temperature, $R$ an effective
radius, $m$ the pion mass, and $\Delta$ a constant with
dimensions of momentum.

Let $\eta_0$ and $\eta $ be the number densities before 
and after symmetrization, respectively. In terms of $g(k,x)$
we have
\be
\eta_0= \int g(k,x)d^4kd^4x
\label{eq:dens}
\ee
and a corresponding expression for $\eta $ with $g$ replaced
by the source function after symmetrization.

Then one finds \cite{Pratt1} that $\eta $ increases with $\eta_0$ and
above a certain crtitical density $\eta^{crit}_0 $, 
$\eta $ diverges. This is
interpreted by Pratt as {\em pasing}.

The reader may be rightly puzzled by the fact that while $\eta $ has a
clear physical meaning the number density
$\eta_0 $ and a fortiori its critical value have no 
physical meaning, because in nature
there does not exist a system of bosons the wf of
which is not symmetrized. That is why we call this approach
``surrealistic".   
 The physical factors which induce condensation 
are, for systems in (local) thermal and chemical equilibrium,
\footnote{For lasers the determining dynamical factor is
among other things the inversion of the occupation of atomic
levels.},
pressure and temperature and the symmetrization is contained
automatically in the form of the distribution function
\be
f=\frac{1}{\exp[(E-\mu]/T]- 1}
\label{eq:f}
\ee 
where 
$E$ is the energy and $\mu$ the chemical potential.

To realize what is going on
 it is useful to observe that the increase of 
$\eta_0 $ can be achieved by decreasing $R$ and/or $T$. Thus
$\eta_0$ can be substituted by one of these two physical
quantities. Then the blow-up of the number density $\eta $
can be thought of as occuring due to a decrease of $T$ and/or
$R$. However this is nothing but the well known
Bose-Einstein condensation phenomenon. 
  
While from a purely mathematical point of view the
condensation effect can be achieved also by starting
with a non-symmetrized wf and symmetrizing it 
afterwards ``by hand" , the causal i.e. physical relationship   
is different: one starts with a bosonic i.e symmetrized
system and obtains condensation by decreasing the
temperature or by increasing the density of this {\em bosonic} 
system. To obtain a pion condensate e.g., the chemical potential
has to equate the pion mass.

A scenario for such an effect in heavy ion reactions has
been proposed in \cite{Ornik}. 

To conclude the ``paser" topic, one must correct 
another confusing interpretation which relates to the 
 observation
made also in \cite{Zajc} that the symmetrization produces a broadening of the
multiplicity distribution (MD). In particular starting with a
Poisson MD for the non-symmetrized wf one ends up
after symmetrization with a negative binomial. While Zajc
correctly considers this as a simple consequence of Bose
statistics, ref.\cite{Pratt1} goes further and associates
this with the so called pasing effect. That such an
interpretation is incorrect is obvious from the fact that
for true lasers
the opposite effect takes place. Before ``condensing" i.e.
below threshold their
MD is in general broad and of negative binomial form corresponding to
a chaotic (thermal) distribution, while above threshold the
laser condensate is produced and as such
corresponds to a coherent state and therefore is
characterized by a Poisson MD.


\begin{thebibliography}{99}
        
\bibitem{Quarkium} P. Carruthers, Collective Phenomena, 
1 (1973) 147-161.
\bibitem{Heretical} P. Carruthers, Annals of The New York 
Academy of Sciences, 229 (1974) 91-123.
\bibitem{LESIP} Local Equilibrium in Strong Interaction
Physics, editors D. Scott and R. Weiner, Bad-Honnef 1984,
World Scientific 1985; Hadronic Matter in Collision, (LESIP II)
editors P. Carruthers and D. Strottman, Santa-Fe 1986, World
Scientific 1986; Hadronic Matter in Collision 1988 (LESIP III)
, editors
P. Carruthers and J. Rafelski, Tucson 1988, World Scientific
1989; Correlations and Multiparticle Production-CAMP (LESIP 
IV), editors M. Pl\"umer, S. Raha, and R. M. Weiner, Marburg
1990, World Scientific 1991.
\bibitem{CZ} P. Carruthers and F. Zachariasen, Rev. Mod.
Phys. 55 (1983) 245.
\bibitem{FW} G.N.Fowler and R.M.Weiner, Phys.
Rev. D 17 (1978) 3118.
\bibitem{int} P. Carruthers, E. M. Friedlander, C. C. Shih,
and R. M. Weiner, Phys. Lett. 222 (1989) 487. 
\bibitem{Pratt} S. Pratt, Phys. Rev. Lett. 53 (1984) 
1219.
\bibitem{bernd1} B. R. Schlei et al., Phys. Lett. B 293 (1992) 275.
\bibitem{Yano} F. B. Yano and S. E. Koonin, Phys. Lett. 78 B
(1978) 556. 
\bibitem{Axel} A. Timmermann et al., Phys. Rev. C 50 (1994) 
3060.
\bibitem{AW} I. V. Andreev and R. M. Weiner, Phys. Lett. B
253 (1991) 416.
\bibitem{APW1} I. Andreev, M. Pl\"umer, R. Weiner, Phys. Rev.
Lett. 67 (1991) 3475. 
\bibitem{2Na49} H. Appelsh\"auser et al., Na49
collaboration, Eur. Phys. J C 2 (1998) 661
\bibitem{Nix} S.Chapman, J.Nix and U.Heinz, Phys. Rev. C 52
(1995) 2694.
\bibitem{APW} I.V.Andr eev, M.Pl\"umer, and R. M. Weiner, 
Int.J.Mod.Phys. A 8 (1993) 4577.
\bibitem{Weiner} R. M. Weiner, Phys. Lett. B 232 (1989) 278;
Phys. Lett. B 242 (1990) 547.
\bibitem{UA1} N. Neumeister et al. (UA1 Collaboration),
Phys.Lett. B 275 (1992) 186. 
\bibitem{mad} M. Pl\"umer, L. V. Razumov, and R. M. Weiner,
Phys. Lett. B 286 (1992) 335.
\bibitem{Coll} Bose-Einstein Correlations in Particle and
Nuclear Physics, A collection of reprints, edited by 
R. M. Weiner, John Wiley and Sons, Chichester, New York, 1997. 
\bibitem{UA1New} H. C. Eggers, P. Lipa, and B. Buschbeck,
 Phys. Rev. Lett. 79 (1997) 197.
\bibitem{intpr} I.V. Andreev et al., Phys.Rev. D 49 (1994)
1217.
\bibitem{bi} A. Bialas, Nucl.Phys. A 545 (1992) 285c; Acta
Phys. Pol. B 23 (1992) 561.
\bibitem{Nelly} N. Arbex, M. Pl\"umer and R. M. Weiner,
Phys. Lett. B 438 (1998) 193.
\bibitem{Na22} N. M. Agababyan et al., Z. Phys. C 68 (1995)
229.
\bibitem{Neu} D. Neuhauser, Phys. Lett. B 182 (1986) 289.
\bibitem{Leo} L. V. Razumov and R. M. Weiner, Phys. Lett. B
319 (1993) 431.
\bibitem{Heinz} C. Slotta , U. Heinz, Phys. Lett. B 391
(1997) 469.
\bibitem{Shuryak} E. V. Shuryak, Sov. J. Nucl. Phys. 18
(1974) 667.
\bibitem{RW} R. M. Weiner, Physics Reports, 
to be published.
\bibitem{Migdal} A. B. Migdal, Rev. Mod. Phys. 50 (1978) 107.
\bibitem{Pratt1} S.Pratt, Phys.Lett. B 301 (1993) 159.
\bibitem{Ornik} U. Ornik, M.Pl\"umer and D. Strottman; Phys.Lett. 
B314(1993)401; U.Ornik et al., Phys.Rev. C 56 (1997) 412.
\bibitem{Shih} G.N.Fowler et al., Phys.Lett. B 253 (1991) 421.
\bibitem{Zajc} W.A.Zajc, Phys.Rev. D 35 (1987) 3396.
\bibitem{Chao} W.Q.Chao, C.S.Gao, and Q.H.Zhang, J.Phys.
G 21 (1994) 847.

\end{thebibliography}
\end{document}